\documentclass[amsmath,amssymb,prb,superscriptaddress,reprint,showpacs,longbibliography]{revtex4-1}
\usepackage{graphicx}
\usepackage{dcolumn}
\usepackage[colorlinks=true,linkcolor=blue,citecolor=blue,urlcolor=blue]{hyperref}
\usepackage[usenames,dvipsnames]{xcolor}
\usepackage{soul}
\usepackage{braket}
\usepackage{bm}
\usepackage{upgreek}
\usepackage{mathtools}  

\begin{document}

\title{Tunneling density of states, correlation energy, and spin polarization \\
 in the fractional quantum Hall regime}

\author{Gaurav Chaudhary}
\email{gaurav-ph@utexas.edu}
\affiliation{Department of Physics, The University of Texas at Austin, Austin, Texas 78712, USA}
\author{Dmitry K. Efimkin}
\affiliation{Department of Physics, The University of Texas at Austin, Austin, Texas 78712, USA}
\affiliation{Department of Physics and Astronomy, Monash University, Clayton, Victoria 3800, Australia}
\author{Allan H. MacDonald}
\affiliation{Department of Physics, The University of Texas at Austin, Austin, Texas 78712, USA}

\date{\today}


\begin{abstract}
We derive exact sum rules that relate the tunneling density-of-states (TDOS) of spinful 
electrons in the fractional quantum Hall (FQH) regime to the spin-dependent many-body ground state correlation energy.
Because the tunneling process is spin-conserving, the 2D (two-dimensional) to 2D
tunneling current $I$ at a given bias voltage $V$ in a spin-polarized system is a sum of majority and minority spin contributions. 
The sum rules can be used to define spin-dependent gaps that we associate with peaks in 
2D to 2D tunneling $I-V$ curves.
We comment in the light of our sum rules on what recent tunneling experiments say about the spin-dependence
of correlation energy contributions, and propose new measurements that could provide more specific experimental estimates.
\end{abstract}

\maketitle

\section{Introduction}
For many years following the discovery of the integer~\cite{Klitzing1980} (IQHE) and fractional~\cite{Tsui1982} (FQHE) quantum Hall effects, the study of two dimensional electron systems (2DES) in a strong perpendicular magnetic field has 
regularly provided examples of distinctly new many-electron physics. Correlations are strong in the FQH regime 
because of kinetic energy quantization, and distinct because of restrictions on correlations imposed by Hilbert space truncation to individual Landau levels. In the limit of strong Landau quantization and weak disorder, electron-electron interactions provide 
the only relevant energy scale.  The set of exotic many electron states discovered in the FQH regime includes incompressible ground states  at a variety of fractional Landau level filling factors $\nu = N/N_{\phi}$ that are 
dramatically signaled by dissipation free edge transport and quantized Hall conductivities. (Here
$N$ is the number of electrons in the system, $N_{\phi}=\Phi/\Phi_0$ is the degeneracy of a Landau level, 
$\Phi$ is the flux through the 2D (two-dimensional) sample, and 
$\Phi_0$ is the electron magnetic flux quantum.)  The elementary charged excitations of incompressible states have fractional charge, and can have non-Abelian statistics~\cite{Read2000} with potential applications in topological quantum computing~\cite{Nayak2008}. In this paper we focus on spin-physics in the FQH regime and on its relationship to bilayer 2D to 2D tunneling.

Although the macroscopic set of degenerate single-particle states within a Landau level can be viewed as an analog of an open atomic shell, the peculiarities of correlations at fractional filling factors often~\cite{Halperine1983,Zhang1984} lead to violations of Hund's rules, {\it i.e.} to incompressible ground states that do not maximize the total spin quantum number.  For example the ground state at $\nu=1$ is maximally spin-polarized whereas the $\nu = 1/2$ ground state is thought to be unpolarized in the absence of Zeeman coupling to an external magnetic field~\cite{Kukushkin1999,Liu2014}.
At $\nu = 2/3,\, 3/5,\, 4/7,\, 2/5,\, 3/7,\, 4/9$ among other filling factors,~\cite{Eisenstein1989,Clarck1989,Kukushkin1999} Zeeman coupling drives first order phase transitions from unpolarized to partial or  fully polarized states, whereas spin-polarization appears~\cite{Kukushkin1999,Liu2014,Eisenstein2016} to increase continuously with Zeeman coupling at the filling factors ($\nu = 1/2, 3/2$) which are mapped to zero-magnetic field by the composite fermion constructions~\cite{Jain1989,Halperin1993}.  
Although there has been considerable progress in understanding the ground state spin-polarization in the FQH regime~\cite{Park2001,Balram2015,Zhang2016}, the role played by spin and Landau level mixing in the experimentally observed double-peaked tunneling at $\nu = 3/2$ versus the single-peaked tunneling at $\nu = 1/2$ is unclear~\cite{Eisenstein2009}.

2D to 2D tunneling experiments in the FQH regime have been an important probe of the correlation physics of underlying FQH states~\cite{Ashoori1990,Ashoori1993,Eisenstein1992} for example by directly measuring Haldane pseudopotentials~\cite{Dial2010,MacDonald2010}. 
These experiments yield non-linear $I-V$ curves with strong current suppresion at low bias voltages~\cite{Ashoori1990,Ashoori1993,Eisenstein1992}.  Early 2D to 2D tunneling experiments were performed mainly in systems with strong enough Zeeman coupling to achieve full spin polarization.  
Recent 2D to 2D tunneling experiments have shifted the  focus to 
$\nu = 1/2, 3/2$~\cite{Eisenstein2009,Eisenstein2016,Eisenstein2018} and 
$\nu = 5/2, 7/2$~\cite{Eisenstein2017} bilayer systems, which have more complicated ground state spin configurations. 
One example of a spin-related surprise is  the presence of double peak structures in tunneling from $\nu = 3/2$ to 
$\nu=3/2$ which are absent at the same Zeeman coupling strength for $\nu = 1/2$ to $\nu=1/2$, suggesting 
partial spin polarization in the $\nu=3/2$ case~\cite{Eisenstein2009,Eisenstein2016}. 
Recent experiments have also shown interesting tunneling characteristics between two FQH layers maintained at different filling by independent gate control~\cite{Eisenstein2017,Eisenstein2019}. 
Additionally, an intriguing ideal spin diode device has been realized using tunneling between FQH layers maintained at 
$\nu = 5/2$ and $\nu = 7/2$, that have different ground state spin polarizations~\cite{Eisenstein2017}.

The goal of this paper is to establish some rigorous sum rules that 
can assist with the interpretation of tunneling I(V) measurements 
between many-electron states in the fractional quantum Hall regime that are in general partially spin-polarized.
Importantly the two fractional quantum Hall states are allowed to have 
different Landau level filling factors.  To this end we extend the tunneling density-of-states (TDOS) sum-rules derived by 
Haussmann {\it et al.}~\cite{Haussmann1996} to the spinful case, and compare the results to features in tunneling $I-V$ curves. 
We also extend the previous analysis to the case 
of tunneling between states with different Landau level filling factors,
emphasizing in the process that the differences between the chemical potentials of the two states 
must be accounted for carefully.  
Using the sum rules we show that important experimental features depend separately on correlations among electrons that have the same spin and among electrons that have opposite spins.  We employ our sum rules to comment 
on the implications of the tunneling measurements by 
Eisenstein {\it et al.}~\cite{Eisenstein2009} for spin-dependent correlation energies in
fractional quantum Hall states.  
We also suggest some similar related measurements
using the uncorrelated quantum Hall states at $\nu = 1$ and $\nu = 2$ as probe layers, 
that could provide even more specific information.

The article is organized as follows. In Sec.~\ref{Sec:Sum_rule_Derivation} we summarize the sum-rule derivation and define important quantities that are related to experimental $I(V)$-curves. In Sec.~\ref{Sec:Analysis} we use our sum-rules to estimate
spin-polarizations and correlation energies of many-body states on the basis of previously published experimental data. 
In Sec.~\ref{Sec:Sum_Rule_Use} we propose some similar tunneling experiments that use quantum states that are 
uncorrelated (in a sense that we will define precisely) as probe layers to extract information about correlations in the 
quantum states of its tunneling partner.  We conclude in Sec.~\ref{sec:discussion}. 
Some details of the derivations are relegated to Appendices~\ref{App:CE_expressions}, \ref{App:Fitting}.


\section{Sum Rules for the Tunneling Density of States \label{Sec:Sum_rule_Derivation}}

We consider a 2DES in strong perpendicular magnetic field in the 
FQH regime where all electrons are in the lowest orbital Landau level (LLL). The single particle states in the symmetric gauge are then labeled by $\bm{m} = (n,\, s)$, where $n$ is angular momentum and $s=\pm1$ labels spin. When projected to the LLL, the single layer Hamiltonian has only interaction ($H_I$) and
Zeeman ($H_Z$) terms:
\begin{align}
     & H = H_I\,+\,H_Z\notag\\
     &\hspace{0.3cm} = \frac{1}{2}  \sum_{n_i,s_i} \; U^{n_1,n_2}_{n_3,n_4}\, c^{\dagger}_{n_1s_1}c^{\dagger}_{n_2s_2}c_{n_3s_1}c_{n_4s_2}\,\notag\\
     &\hspace{2cm}-\,\lambda_z \sum_{n,s}\, c^{\dagger}_{ns}c_{ns} s\, . 
\label{Eq:Hamiltonian}
\end{align}

Here $c^{\dagger}_{ns}$ $(c_{ns})$ creates (annihilates) an 
electron with angular momentum $n$ and spin $s$, $U$ is the electron-electron interaction such that the interaction matrix element $U^{\bm{m}_1,\bm{m}_2}_{\bm{m}_3,\bm{m}_4} = U^{n_1,n_2}_{n_3,n_4}\delta_{s_1,s_3}\delta_{s_2,s_4}$, $\lambda_z = g\mu_BB/2$ is the Zeeman coupling strength, $g$ is the electron $g$-factor, $\mu_B$ is the Bohr magneton, and $B$ is the magnetic field. 
Because the interaction Hamiltonian is spin-independent, the component of spin along the magnetic field direction and the number of electrons with a given spin label $N_s$ are both good quantum numbers.

We allow for independent control over filling factors of the two layers and assume that the 
individual layers of the bilayer are sufficiently far apart that we can neglect interlayer interactions.  
When interlayer interactions are weak, they yield a small excitonic correction to the results we obtain below that is 
discussed briefly in Section~\ref{Sec:Sum_Rule_Use} and \ref{sec:discussion}. 
The individual layers are then 
coupled only by the single-particle interlayer tunneling term.  For a translationally invariant tunnel barrier and 
temperature $T=0$, the spin-$s$ interlayer current $I$ at the lowest order in the tunneling 
amplitude $t_0$ is~\cite{Mahan2000,Haussmann1996}
\begin{align}
    & I_s(V) = I_0\,
    \int^{eV}_{0}d\epsilon\,  \tilde{A}^{+}_s(\epsilon)\,\tilde{A}^{-}_s(\epsilon-eV)\,d\epsilon\, .
\label{Eq:IV_Convolution}    
\end{align} 
Here the constant $I_0 = et^2_0 S_0/(\hbar\ell^2)$, $S_0$ is the area of 2D system, $\ell$ is the magnetic length, $V >0$ is the applied bias between the two layers, and $\tilde{A}^+_{s}(\epsilon)$ and $\tilde{A}^-_{s}(\epsilon)$ are respectively spin-$s$ electron and 
hole spectral functions of the two individual layers measured from their chemical potentials $\mu$.
The simple form for the $I(V)$ curve reflects the property that the spectral function is the same for every state within a 
Landau level, a property that follows from translational invariance.  The spectral function is however spin $s$ dependent unless
the ground state total spin quantum number is $S=0$. 
We choose the spectral function normalization convention in which the integral over energy of 
$\tilde{A}_{s}(\epsilon)  \equiv \tilde{A}^+_{s}(\epsilon) + \tilde{A}^-_{s}(\epsilon)$ is equal to one.
$\tilde{A}^+_{s}(\epsilon)$ is non-zero only for $\epsilon > 0 $ and only if spin-$s$ is not completely full,
whereas $\tilde{A}^-_{s}(\epsilon)$ is non-zero only for $\epsilon < 0$ and only if spin-$s$ is partially full.
With this understanding the limits on the interval of integration in Eq.~\ref{Eq:IV_Convolution} 
are superfluous, and we can view the integral as a convolution.

The exact microscopic expression for the spectral function is 
\begin{subequations}\label{Eq:TDOS} 
\begin{align}
        & A_s(\epsilon) = A^{+}_s(\epsilon) + A^{-}_s(\epsilon)\, ,\\
        & A^{+}_{s}(\epsilon) = \sum_{\alpha}|\langle \Psi_{\alpha}(N+ 1)|\,c^{\dagger}_{ns}\,|\Psi_0(N)\rangle|^2\notag\\
    &\hspace{2cm}\times\delta(\epsilon-[E_{\alpha}(N+1)-E_0(N)])\, ,\\
    & A^{-}_{s}(\epsilon) = \sum_{\alpha}|\langle \Psi_{\alpha}(N-1)|\,c_{ns}\,|\Psi_0(N)\rangle|^2\notag\\
    &\hspace{2cm}\times\delta(\epsilon-[E_0(N)-E_{\alpha}(N-1)])\,.   
\end{align}
\end{subequations}
Here $E_{\alpha}(N)$ and $|\Psi_{\alpha}(N)\rangle $ are $N$-electron eigenenergies and eigenvectors, 
and $\alpha=0$ denotes the ground state.  Note that in these expressions energy is measured from some physical reference value,
in our case from the energy of the degenerate Landau level, rather than from the chemical potential.
To make this distinction clear we consistently use a $\tilde{.}$ accent for the spectral functions 
with its energy argument measured from the chemical potential and drop
the accent when energies are measured relative to a fixed energy. 
The chemical potential $\mu = \partial E/\partial N$ for these strongly 
correlated electrons depends non-trivially on filling and correlation and at $T=0$ is not pinned by bulk physics 
when the filling factor is exactly equal to an incompressible value.
$A^{+}_s(\epsilon)$ is only non-zero for $\epsilon \geq \mu$, and $A^{-}_s(\epsilon)$ is only non-zero for $\epsilon \leq \mu$.
Below we refer to $A_{s}(\epsilon)$ as the tunneling density-of-states (TDOS).  
For the general case of tunneling between FQH layers maintained at different filling factors, the distinction 
between spectral functions with energy measured from the Landau level energy and 
energy measured from the chemical potential plays an important role.

The TDOSs defined above encode information about ground state correlation energies and for a general strongly correlated system are not known exactly, which makes it difficult to relate experimental $I-V$ curves to microscopic energies of the system. The sum rules are simple expressions for energy moments of both the particle-removal portion of the TDOS, which lies below the chemical potential $\mu$, and the particle-addition portion of the TDOS which lies above the chemical potential. In what follows, we will show that these sum rules of moments of TDOS, which have an exact expressions as function of filling and ground state correlation energies as evaluated below, can be related to moments of $I-V$ curves, which is obtained from experimental data. This way it helps extract the important ground state properties of strongly correlated FQH states from tunneling data. We denote the moments of TDOS by $M^{\gamma,i}_s$, where $\gamma$ is the order of the moment, $i = \pm$ refers to electron addition or removal respectively, and $s$ refers to spin. For the zeroth moment a standard calculation yields
\begin{widetext}
\begin{subequations}
\label{Eq:Sum_zero} 
\begin{align}
     & M^{0,+}_s = \int^{\infty}_{\mu} d\epsilon\,A^{+}_s(\epsilon) = \langle \Psi_0(N)|\, c_{ns}\,\biggl (\sum_{\alpha} |\Psi_{\alpha}(N+1)\rangle\langle \Psi_{\alpha}(N+1)|\biggr )\,c^{\dagger}_{ns}\,|\Psi_{0}(N)\rangle = \bar{\nu}_s\, ,\\
     & M^{0,-}_s = \int^{\mu}_{-\infty} d\epsilon\,A^{-}_s(\epsilon) = \langle \Psi_0(N)|\, c^{\dagger}_{ns}\,\biggl (\sum_{\alpha} |\Psi_{\alpha}(N-1)\rangle\langle \Psi_{\alpha}(N-1)|\biggr )\,c_{ns}\,|\Psi_{0}(N)\rangle = \nu_s\, .    
\end{align}
\end{subequations}
\end{widetext}
Here $\bar{\nu}_s = 1-\nu_s$. This result is similar to the spinless case except that the Landau level filling factor $\nu$ is replaced by the spin-dependent partial filling factor $\nu_s=N_s/N_{\phi}$.

To derive additional sum-rules we consider the equation of motion (EOM) of the time-ordered Greens function,
\begin{equation} 
{G}_s(t) = i\,\langle \langle\, \mathcal{T}[c_{(n,s)}(t)\,c^{\dagger}_{(n,s)}]\,\rangle \rangle\, .
\label{TOGF}
\end{equation} 

In Eq.~\ref{TOGF} $\mathcal{T} [c_{ns}(t) c^{\dagger}_{ns}] \equiv
\theta(t)\,c_{ns}(t)\,c^{\dagger}_{ns}-\theta(-t)\, c^{\dagger}_{ns}\,c_{ns}(t)$ and the double angle brackets imply quantum thermal averages. It follows that 
\begin{align}
    & i\dot{G}_s(t) = i\theta(t)\,\langle \langle\, [H,c_{ns}(t)]\,c^{\dagger}_{ns}\,\rangle \rangle\notag\\
    &\hspace{1.5cm}-i\theta(-t)\,\langle \langle\, c^{\dagger}_{ns}\,[H,c_{ns}(t)]\,\rangle \rangle+\delta(t) .
\label{Eq:GF_EOM}
\end{align}

We evaluate the RHS of  Eq.~\ref{Eq:GF_EOM} at time $t=0^{\pm}$

\begin{widetext}
\begin{align}
    &  \langle \langle\, [H_I,\,c_{ns}(t=0^+)]\,c^{\dagger}_{ns}\rangle\rangle =  \frac{1}{2} \langle\langle \sum_{n_i,s_i} U^{n_1,n_2}_{n_3,n_4}\, (c^{\dagger}_{n_1s_1}c^{\dagger}_{n_2s_2}c_{n_3s_1}c_{n_4s_2}c_{ns}-c_{ns} c^{\dagger}_{n_1s_1}c^{\dagger}_{n_2s_2}c_{n_3s_1}c_{n_4s_2})\,c^{\dagger}_{ns}\,\rangle \rangle
    \notag\\
   & \hspace{3.75cm}= \frac{1}{2}\langle \langle \sum_{\substack{n_1,n_3,n_4\\s_1,s_3,s_4}}
   (U^{n_1,n}_{n_3,n_4}-U^{n,n_1}_{n_3,n_4} )\notag\\
   &\hspace{5.0cm}\times (c^{\dagger}_{n_1s_1}c^{\dagger}_{ns}c_{n_3s_1}c_{n_4s_2}+ c^{\dagger}_{n_1s_1}c_{n_3s_1}\delta_{nn_4}\delta_{ss_2}- c^{\dagger}_{n_1s_1}c_{n_4s_2}\delta_{nn_3}\delta_{ss_1})\,\rangle\rangle\, .
   \label{moment}
\end{align}

\end{widetext}

Since translational invariance guarantees that both sides of Eq.~\ref{moment} are independent of $n$, we can average over this variable to obtain 
\begin{align}
    & \langle \langle\, [H_I,\,c_{ns}(0^+)]\,c^{\dagger}_{ns}\rangle\rangle = 2\, (\epsilon_{s,s} \,+\, \epsilon_{s,-s}\,-\,\nu_s\epsilon_0)\, .
\label{Eq:Interaction_EOM_add}    
\end{align}

Here $\epsilon_0 = E_0(N=N_\phi)/N_{\phi} = -\sqrt{\pi}e^2/(\sqrt{8}\kappa\ell)$ is the energy of a filled Landau level ($\kappa$ is the dielectric constant) and $\epsilon_{s,s}$ and $\epsilon_{s,-s}$ are respectively the contributions to the interaction energy per flux quantum due to interactions among electrons with spin $s$, and due to interactions between electrons with spin $s$ and those with the opposite spin $-s$ (For explicit expressions see Eqs.~\ref{Eq:SM_Correlation_same}, \ref{Eq:SM_Correlation_opposite} in the Appendix~\ref{App:CE_expressions}). Both $\epsilon_{s,s}$ and $\epsilon_{s,-s}$ depend in a complex way on 
minority and majority spin filling factors. A similar calculation leads to 
\begin{align}
     & \langle \langle\, c^{\dagger}_{ns}\, [H_I,\,c_{ns}(0^-)]\,\rangle\rangle = 2\,(\epsilon_{s,s}\, + \,\epsilon_{s,-s})\, .
\label{Eq:Interaction_EOM_sub} 
\end{align}

The contributions to these commutators from the Zeeman term in the Hamiltonian is straightforward to evaluate: 
\begin{subequations}\label{Eq:Zeeman_EOM}   
\begin{align}
   &\langle\langle\,[H_Z, c_{ns}(0^+)]\,c^{\dagger}_{ns}]\,\rangle\rangle 
    = s\bar{\nu}_s\lambda_z \, ,\\
    &\langle\langle\,c^{\dagger}_{ns}[H_Z, c_{ns}(0^-)]]\,\rangle\rangle =
    -s\nu_s\lambda_z\, . 
\end{align}
\end{subequations}

Inserting Eqs.~\ref{Eq:Interaction_EOM_add}, \ref{Eq:Interaction_EOM_sub}, \ref{Eq:Zeeman_EOM} in Eq.~\ref{Eq:GF_EOM} we find that 
\begin{subequations}\label{Eq:GF_EOM_Corr_Energy} 
\begin{align}
    & i\dot{G}_s(0^+) = 2i\,[\,\epsilon_{s,s}+\epsilon_{s,-s}-\nu_s\epsilon_0+\frac{s}{2}\bar{\nu}_s\lambda_z \,]\, ,\\
    & i\dot{G}_s(0^-) = 2i\,[\,\epsilon_{s,s}+\epsilon_{s,-s}-\frac{s}{2}\nu_s\lambda_z \,]\, .
\end{align}
\end{subequations}

The LHS of Eq.~\ref{Eq:GF_EOM_Corr_Energy} can be related to the TDOSs using the spectral representation of the Green's function: 
\begin{align}
    & G_s(\epsilon) = \int^{\infty}_{-\infty}\, dt\, G_s(t)\,e^{i\epsilon t}\notag\\
    & \hspace{0.5cm} = \int^{\infty}_{\mu}\,d\epsilon'\, \frac{A^{+}_s(\epsilon')}{\epsilon-\epsilon'+i\eta}\,+\,
    \int^{\mu}_{-\infty}\,d\epsilon'\, \frac{A^{-}_s(\epsilon')}{\epsilon-\epsilon'-i\eta}\, .
\label{Eq:GF_Spectral}    
\end{align}

Since 
\begin{align}
    & i\dot{G}_s (t) = \int^{\infty}_{-\infty}\,\frac{d\epsilon}{2\pi}\,\epsilon
    \,G_s(\epsilon)\,e^{-i\epsilon t}\, ,
\label{Eq:GF_EOF_Spectral}    
\end{align}
first moment sum rules follow from Eqs.~\ref{Eq:GF_EOM_Corr_Energy}, \ref{Eq:GF_Spectral} , and \ref{Eq:GF_EOF_Spectral}
\begin{subequations}\label{Eq:Sum_first}
\begin{align}
   &  M^{1,+}_{s} = \int^{\infty}_{\mu}\,\epsilon A^+_s(\epsilon)\,d\epsilon\notag\\  
   &\hspace{0.8cm}= -2\epsilon_{s,s}-2\epsilon_{s,-s}+2\nu_{s}\epsilon_{0}
    -s\bar{\nu}_s\lambda_z\,   ,\\
    & M^{1,-}_{s} = \int^{\mu}_{-\infty}\, \epsilon A^{-}_s(\epsilon)\,d\epsilon\notag\\ 
    &\hspace{0.8cm}= 2\epsilon_{s,s}+2\epsilon_{s,-s}-s\nu_s\lambda_z\, .    
\end{align}
\end{subequations}
In contrast to the spinless case, the first order moment sum-rules for the spinful case depend on a partitioning of ground state energy contributions based on spin.

Because of the long-range of the Coulomb interaction between electrons, the electrostatic energy 
contribution to the electron energies above are infinite.  For an isolated uniform 
density this contribution to the energy can be easily accounted for by taking the zero of 
energy at $-e\varphi_{es}$, where $\varphi_{es}$ is the electrostatic potential evaluated in the  layer
contributed by both electrons and neutralizing positive charges, which are in practice normally located on gates.
The energies are then understood to exclude the mean-field electrostatic contribution.
To interpret tunneling experiments between different layers, we have to 
be careful to keep track of how we choose their zeroes of energy.  To relate to tunneling experiments,
the most convenient choice is to choose the zero of energy at the local chemical potential of each layer.
We therefore defined an alternate set of moment 
sum-rules for the spectral functions defined in this way:
 \begin{subequations}\label{Eq:zero_moment_fermi}
\begin{align}
     &\tilde{M}^{0,+}_s = \int^{\infty}_{0} d\epsilon\, \tilde{A}^{+}_s(\epsilon) = \bar{\nu}_s\, , \\
     &\tilde{M}^{0,-}_s = \int^{0}_{-\infty} d\epsilon\, \tilde{A}^{-}_s(\epsilon) = \nu_s\, ,
\end{align} 
\end{subequations}
and,
\begin{subequations}\label{Eq:1st_moment_fermi}
\begin{align}
     &\tilde{M}^{1,+}_s\, =\, \int^{\infty}_{0} \epsilon \tilde{A}^{+}_s(\epsilon)\,d\epsilon\, =\, \int^{\infty}_{\mu} (\epsilon-\mu)A^{+}_s(\epsilon)\,d\epsilon\,\notag\\
     &\hspace{0.9cm} = \,M^{1,+}_s-\mu\bar{\nu}_s\, , \\
     &\tilde{M}^{1,-}_s\, =\, \int^{0}_{-\infty} \epsilon \tilde{A}^{-}_s(\epsilon)\,d\epsilon\, =\, \int^{\mu}_{-\infty}(\epsilon-\mu)A^{-}_s(\epsilon)\,d\epsilon\,\notag\\
     &\hspace{0.9cm} =\, M^{1,-}_s-\mu\nu_s\, .
\end{align} 
\end{subequations}

We see that while the zeroth order sum-rules are unchanged by a shift in the zero of energy,
the first order sum rules for the spectral functions measured from chemical potential have an additional correction, which has not  been considered in earlier work~\cite{Haussmann1996}.
Since the tunneling $I-V$ peaks are related to differences between moment ratios (see below) 
the chemical potential terms play a role only when the chemical potentials of the layers are different, 
either because the states are at different filling factors or because they have different electrostatic 
potentials.  For tunneling between identical states the results obtained by Haussmann \textit{et al.}~\cite{Haussmann1996} remain valid.  As we emphasize below, however, the 
chemical potential corrections are important for tunneling between states  
with densities slightly on opposite sides of incompressible filling factors, which should not be considered identical.

We now define an effective spin-dependent gap by calculating the difference between the
average energy of electrons added to the system and electrons removed from the system. 
This gap is intended for comparison with the voltage bias at peak current in the bilayer tunneling experiments: 
\begin{align}
    &\Delta_s = \frac{\tilde{M}^{1,+}_s}{\tilde{M}^{0,+}_s}\biggr |_t\,-\,\frac{\tilde{M}^{1,-}_s}{\tilde{M}^{0,-}_s}\biggr |_b\notag\\
    &\hspace{0.4cm} = 2\epsilon_0\frac{\nu^{t}_s}{\bar{\nu}^{t}_s} - \frac{2}{\nu^{b}_s}(\epsilon^{b}_{s,s}+\epsilon^b_{s,-s}) - \frac{2}{\bar{\nu}^t_s}(\epsilon^{t}_{s,s}+\epsilon^{t}_{s,-s})\notag\\
    &\hspace{0.8cm}+(\mu^{b}-\mu^t)\, .
\label{Eq:Gap}    
\end{align}
The indices $t, b$ above stand for top and bottom layer, and layer $b$ is assumed to 
have a higher chemical potential. This assumption 
allows us to replace the $\pm$ indices by $t$ and $b$ indices, keeping in mind that 
the electron addition spectral function is always associated with top
layer and electron removal with the bottom layer. 
In our interpretation, $I-V$ curves with two peaks are strongly suggestive 
of spin-dependent energy gaps $\Delta_s$.
A more informative expression for the gap can be obtained by separating the 
total interaction energy of the two-dimensional electron gas into exchange and 
correlation contributions using: 
\begin{equation}
\epsilon_{s',s} = \delta_{s',s} \nu_s^2 \epsilon_0 + \epsilon^c_{s',s}.
\label{eq:xplusc}
\end{equation}
The first term on the RHS of Eq.~\ref{eq:xplusc} is the interaction average 
energy of all states in the single-Landau level Hilbert space and is obtained when 
single-particle states in the lowest Landau level Hilbert space are occupied randomly, which  
is negative in the presence of a neutralizing background simply because of electron-avoidance
due to the Pauli exclusion principle.  The correlation energy $\epsilon^c_{s',s}$ is particle-hole symmetric
in the lowest Landau level Hilbert space and represents the additional energy gained when particles avoid 
interactions to the maximum degree allowed by the Hilbert space constraint.  
For the fully spin-polarized Laughlin states at $\nu=1/3$, for example, 
the exchange energy per flux is $-0.07 \, e^2/\kappa \ell$ whereas the correlation energy per flux 
is $-0.067 \, e^2/\kappa \ell$~\cite{Fano1988}.  The particle-hole counterpart of this state in the 
$n=0$ Hilbert space has $\nu=5/3$ and using Eq.~\ref{eq:xplusc} along with the interpolation from Fano \textit{et. al.}~\cite{Fano1988}, the exchange energy per flux is $-0.905 \, e^2/\kappa \ell$,
but the same correlation energy.
When expressed in terms of correlation energies the spin-dependent gap 
has the form: 
\begin{align}
    &\Delta_s = 2 \epsilon_0 (\nu^{t}_s-\nu^{b}_s) + (\mu^{b} - \mu^{t})\notag\\   
    &\hspace{1cm}- \frac{2}{\nu^b_s}(\epsilon^{cb}_{s,s}+\epsilon^{cb}_{s,-s})-\frac{2}{\bar{\nu}^t_s}(\epsilon^{ct}_{s,s}+\epsilon^{ct}_{s,-s})\, .
\label{Eq:Gapc}    
\end{align}
The first term on the RHS of Eq.~\ref{Eq:Gapc} 
is the difference of the exchange self-energies of the two layers.  Note that this term cancels the 
exchange contribution to the chemical potential difference between layers, the second term in Eq.~\ref{Eq:Gapc}.
It follows that the gap vanishes when correlations are neglected.  This is consistent with the 
fact that when interactions are treated at the Hartree-Fock level, 
the TDOSs are $\delta$-functions in energy that are pinned to the chemical potential.

Having established separate sum rules for the electron addition and electron removal contributions to the spin resolved spectral functions, we now write down corresponding tunneling currents sum rules:
\begin{align}
    & P^0_s = \int^{\infty}_0 \,I_s(\epsilon)\,d\epsilon = I_0 \int^{\infty}_{0}\,\tilde{A}^t_s(\epsilon)\,d\epsilon \int^{0}_{-\infty}\,\tilde{A}^b_s(\epsilon)\,d\epsilon \notag\\
    &\hspace{2.7cm} = I_0\bar{\nu}^t_s\nu^{b}_s\, .
\label{Eq:Integral_IV}    
\end{align}
and 
\begin{align}
    & P^{1}_s = \int^{\infty}_0\, \epsilon\, I_s(\epsilon)\, d\epsilon\notag\\
    & \hspace{0.4cm}= P^0_s \,\Delta_s\, .
\label{Eq:current_moment_1st}      
\end{align}
The second line of Eq.~\ref{Eq:current_moment_1st} is valid when  interlayer interactions are negligible.
Below we assume that interlayer interaction effects are weak and that any excitonic shifts they yield in
the weak interaction limit have been corrected for before our sum rules are applied.  
Note that the ratio of first and zeroth order moment of the spin resolved current is equal to the spin dependent gap defined in Eq.~\ref{Eq:Gap}.


\section{Sum Rules and Correlation Energies \label{Sec:Analysis}}

In this section we illustrate how our sum-rules can shed light on spin-dependent 
correlations by using them to interpret $I-V$ data obtained in tunneling studies 
of partially spin-polarized FQH states and the role the correction due to chemical potential difference. 
We separately discuss the compressible case of $\nu = 1/2$ and $\nu = 3/2$ from $I-V$ curves obtained by Eisenstein {\it et al.}~\cite{Eisenstein2009} in thin quantum well samples and a general incompressible case. 
As we will discuss, the extraction of spin-information from 
current data depends on  curve-fitting that has some uncertainty.
Nevertheless the conclusions we reach are sensible and interesting.
In the following section 
we discuss additional measurements that could
provide more reliable spin-dependent correlation energies partial filling
factors.

\subsection{Tunneling at compressible filling factors}
\begin{figure}[!htb]
  \includegraphics[width=0.45\textwidth]{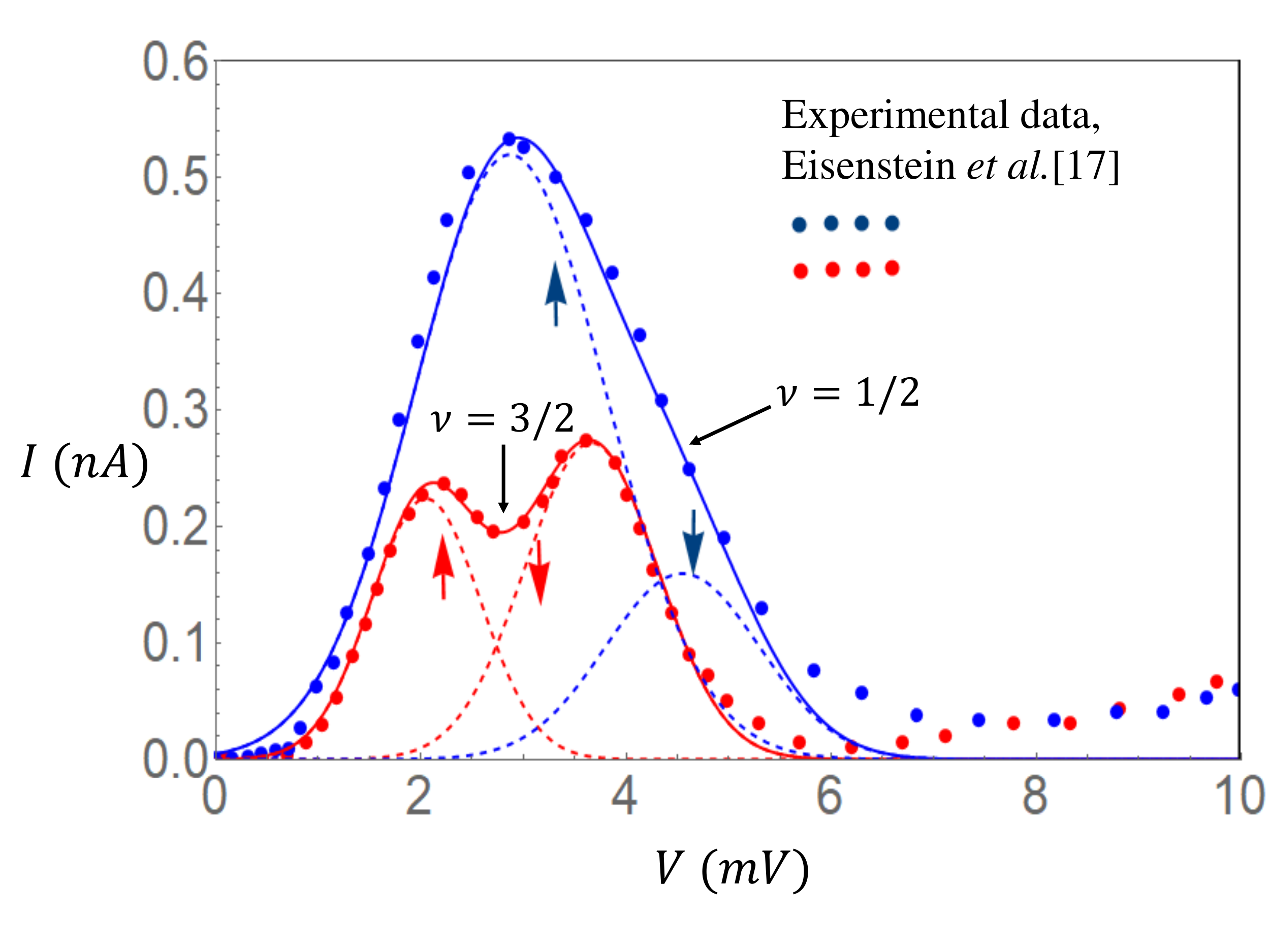}
  \caption{\label{Fig:Eisenstein_fit}
Three-parameter fits (solid line) to the experimental $I-V$ data in Ref.\onlinecite{Eisenstein2009} assuming that the 
contributions from each spin-component can be approximated by a Gaussian specified by a peak voltage, a 
strength and a width. The dashed curves show the spin resolution of the fit. ($I\propto\sum_s \exp(-(V-\Delta_s)^2/2\sigma_s^2)$  Voltages are expressed in energy units. Gaussian currents correspond to Gaussian electron removal and 
electron addition spectral function contributions provided that the spectral weight at the Fermi level is negligible.
We find that for $\nu=3/2$ partial fillings are $\nu_{\uparrow} = 0.818$ and $\nu_{\downarrow} = 0.682$, while for $\nu=1/2$ partial fillings are  $\nu_{\uparrow} = 0.35$ and $\nu_{\downarrow} = 0.15$  
}
\end{figure}

In the limit of negligible Landau level mixing, the tunneling $I-V$'s at $\nu=1/2$ and $\nu=3/2$ are required to be identical by particle-hole symmetry within the spinful lowest orbital LL Hilbert space.  
Experiment provides clear evidence of particle-hole symmetry breaking that is 
evident, for example, in the much larger ~\cite{Kukushkin1999,Eisenstein2009,Liu2014} critical Zeeman energy at which spin-polarization becomes complete in the $\nu = 3/2$ case.  The importance of Landau level mixing at a given carrier density and filling factor is not universal, but depends on sample thickness~\cite{Liu2014} among other details. Below we assume that all LL mixing effects can be approximated 
by changes in the effective interactions between electrons~\cite{Sodemann2013} so that our sum rules apply. 
Our analysis also assumes weak interlayer interactions as mentioned previously.

The double-peak structure of the $\nu = 3/2$ $I-V$ curve suggests that majority and 
minority spin electrons both make substantial contributions to the tunneling current 
and the difference between electron addition and removal energies is spin-dependent.
If the electron addition and removal contributions to the spin-resolved spectral functions 
can be approximated by Gaussians, the $I-V$ curve is a convolution (Eq.~\ref{Eq:IV_Convolution}) of Gaussians, and therefore also Gaussian.  The experimental $I-V$ curve is not strictly Gaussian of course, since 
it must vanish at zero bias voltage and is known to be strong suppressed at low bias 
due to the Coulomb gap effect~\cite{Ashoori1990, Eisenstein1992, Eisenstein2016}  and skewed at high bias,
possibly due to the influence of disorder. 
Although the Coulomb gap at low bias has some very interesting~\cite{Chowdhury2018,Chowdhury2018a} 
features also seen in experiment~\cite{Eisenstein2016}, the sum rule moments on which we 
focus are dominated by the behavior of the $I-V$ curve near its peak, and are largely uninfluenced by low 
bias behavior. 
We attempt to extract physics from the $I-V$ curves by fitting them to the equation 
\begin{equation}
I(V) = I_0 \sum_s  \frac{\nu_s \bar{\nu}_s}{\sqrt{2}\pi\sigma_s} \exp[-(V-\Delta_s)^2/2\sigma^2_s].
\label{eq:fitI}
\end{equation}
Here $I_0$ is not normally known accurately.  The factors to $\bar{\nu}_s$ and $\nu_s$ in Eq.~\ref{eq:fitI} 
are motivated by the zeroth order sum rules, and the peaks for the individual 
Gaussians are associated with the spin-dependent gaps. For $\nu = 3/2$ good fits can be obtained by setting $\Delta_s$ to the two peak biases and, and fixing the $\sigma_s$ values to describe the two peak widths.   
Lastly the relative peak height is adjusted to determine the spin-dependent 
partial filling factors. (See Eq.~\ref{Eq:Sup_filling_ratio} in the Appendix~\ref{App:Fitting}). 
For $\nu = 1/2$ fitting, a single Gaussian associated with majority spin is first assigned with the bias at the peak current. Then $\sigma_{\uparrow}$ is determined by peak width towards the lower bias side of the peak current. Lastly another relatively small Gaussian is added at higher bias to match the skewness in the experimental data (See the Appendix~\ref{App:Fitting} for more detail on the fitting procedures).  Here and subsequently we use $\uparrow$ and $\downarrow$ to denote majority and minority spins respectively.

For $\nu=3/2$, we conclude that the partial filling factors are $\nu_{\uparrow} = 0.818$ and $\nu_{\downarrow} = 0.682$, implying about $9\%$ polarization, compared to the $33\%$ maximal 
spin polarization at this filling factor. 
$\nu = 1/2$ does not show a clear double peak structure, although it is 
expected to be partially spin-polarized~\cite{Eisenstein2016}, and the spin-decomposition 
is less certain. We associate the main peak with the majority spin, 
and associate the skewness at higher bias with a weak 
minority spin contribution.  In our fit $\nu_{\uparrow}\sim 0.35$ and $\nu_{\downarrow}\sim 0.15$ 
giving about $40\%$ polarization.

From the spin-dependent gaps, spin-dependent correlation energies 
can be estimated using Eq.~\ref{Eq:Gapc}.  We conclude that
\begin{subequations} 
\begin{align}
    \begin{rcases}
     &\epsilon^c_{\uparrow\uparrow}+\epsilon^c_{\uparrow\downarrow} \sim -0.017e^2/\kappa\ell \\
     &\epsilon^c_{\downarrow\downarrow}+\epsilon^c_{\uparrow\downarrow} \sim  -0.046e^2/\kappa\ell 
     \end{rcases}\text{$\nu = 3/2$}\, \\
     \begin{rcases}
     & \epsilon_{\uparrow\uparrow}^c+\epsilon_{\uparrow\downarrow}^c \sim -0.037e^2/\kappa\ell\\
      &\epsilon_{\downarrow\downarrow}^c+\epsilon_{\uparrow\downarrow}^c \sim -0.033e^2/\kappa\ell
     \end{rcases}\text{$\nu = 1/2$}\, .
\end{align}
\end{subequations}
Here energies were converted into the standard  $e^2/\kappa\ell$ units of fractional quantum
Hall systems using the dielectric constant $\kappa = 12.9$ of GaAs and the magnetic field at which these 
experiments were performed.  The left hand sides of these equations can be viewed as total correlation energies of electrons of a given spin due to interactions with  
other electrons of the same spin and electrons of the opposite spin:
$\epsilon^c_{s} \equiv \epsilon^c_{s,s}+\epsilon^c_{s,-s}$.
Note that there is no exchange energy contribution to the gap in this case because the filling factors 
on opposite sides of the tunnel barrier are equal.
As expected the total correlations energies are similar in the two cases because the mobile 
electron carriers at $\nu=1/2$ have the same density as the mobile hole carriers at $\nu=3/2$.  
For $\nu=3/2$ minority spins dominate the correlation energy because they have a higher hole 
density, whereas for the more weakly polarized $\nu=1/2$ state the majority spins have a 
larger correlation energy as expected.  
Although the sum rules do provide an estimate for the difference between the correlation 
energy contributed by interactions between majority and minority spins, the determination remains 
somewhat uncertain due to the vagaries of the fitting procedure.  
In the next section we explain how new types of tunneling measurements could be used to determine these quantities 
uniquely.

\subsection{Tunneling at incompressible filling factors}

\begin{figure*}[!htb]
  \includegraphics[width=1.0\textwidth]{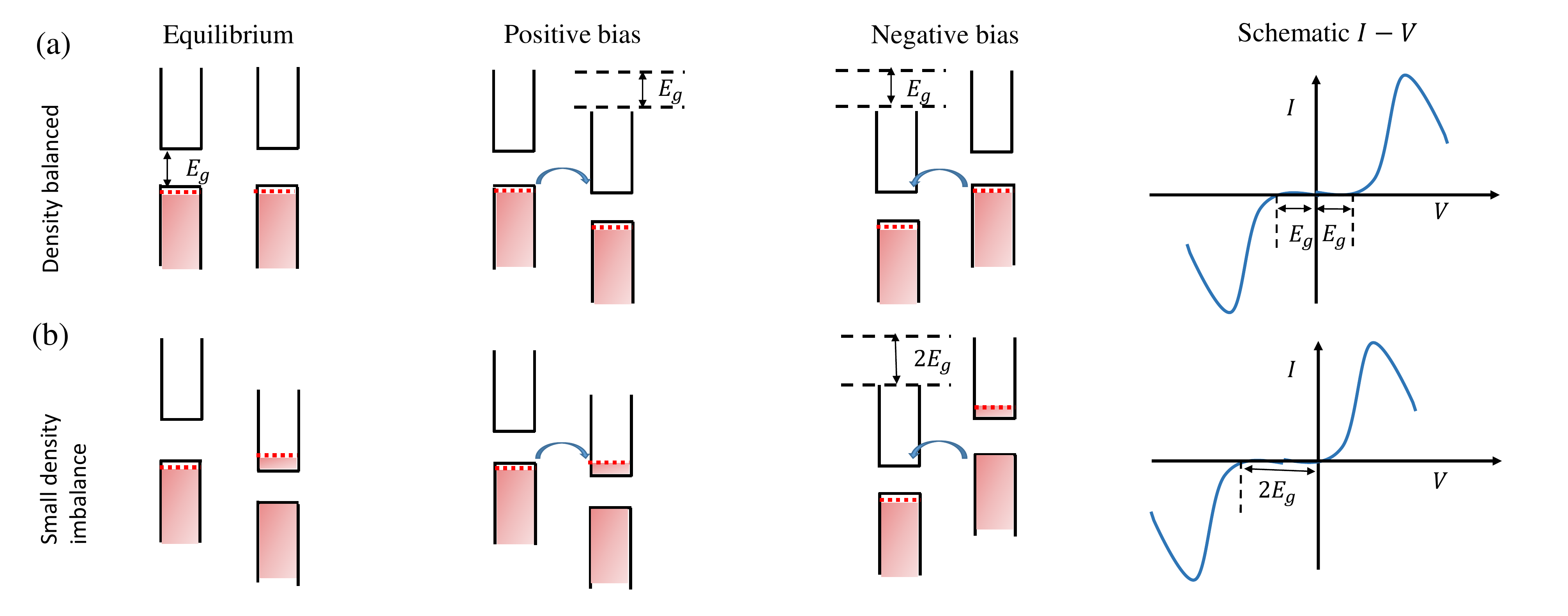}
  \caption{\label{Fig:Incompressible_schematic}
  Schematic of tunneling between two quantum wells that are close to incompressible FQH filling factors. 
  The blue arrows represent direction of electron flow.
  (a) When both layers are on the same side of the incompressible filling factor
  the spectral functions line up in equilibrium. 
  The tunneling $I-V$ has a hard gap equal to the chemical potential gap $E_g$ for both signs of bias. 
  (b) When the equilibrium densities in the two layers are on opposite sides of an incompressible value,
  the spectral functions are offset by $E_g$ in equilibrium.  The offsets shift the hard 
  transport gaps by $\pm E_g$ depending on the sign of the bias voltage to $0$ and $2 E_g$.  
For one sign of bias, there is negligible gap, while for the other sign of bias, the current is suppressed for biases up to 
$2E_g$.}
\end{figure*}

The dependence of the tunneling $I-V$ on density 
near incompressible filling factors can be used to extract estimates of the 
chemical potential jumps. 
To illustrate this point, consider tunneling between layers that are close to the same filling factor,:
$\nu = \nu_i\pm\delta\nu$. Here $\nu_i$ denotes some incompressible FQH filling factor. 
When the layers are both slightly below or slightly above the incompressible filling $\nu_i$,
tunneling can begin at bias voltage magnitudes roughly equal to the chemical potential gap $E_g$ for either bias voltage sign,
as shown in Fig.~\ref{Fig:Incompressible_schematic}(a).  Correlation effects can of course 
substantially suppress tunneling current even above this bias voltage magnitude if $E_g$ is small compared 
to the spectral function gap $\Delta$.  
When the filling factors in the two layers are on opposite sides of the incompressible filling factor
the tunneling curve is not an odd function of bias voltage, as shown in Fig.~\ref{Fig:Incompressible_schematic}(b).  
The chemical potential jump upon crossing the gap shifts the tunneling $I-V$ by $E_g$ in opposite 
directions for opposite signs of bias voltage.  These shifts allow the chemical potential 
gaps to be measured by 
performing a tunneling 
experiment, even though contributions to the spectral function at energies of the order of transport gap are 
very small and not directly measurable in fractional quantum Hall systems because of their 
strong electronic correlations.


\section{Filled Landau Levels as Spin Probes \label{Sec:Sum_Rule_Use}}

As we have explained, 2D to 2D tunneling experiments are sensitive probes of the 
ground state correlation energies of FQH states and reveal important details about the nature of the state.  
In the non-linear tunneling $I-V$ curves, a strong suppression appears near small bias voltages $V$, which is 
a common characteristic of strongly correlated electron states sometimes referred to as the Coulomb gap.
The energy required to add an electron is on average larger than the energy gained by removing an 
electron by a finite value independent of whether or not the system is incompressible, {\it i.e.} independent of whether or not it has 
a strict gap for the lowest energy charged excitations. 

In this section we explain how 
the Slater-determinant many-body ground states  
of fully spin-polarized $\nu=1$ states and unpolarized $\nu = 2$ states 
can be used together as a very specific tunnel-probe of the spin-dependent 
correlations in a non-trivial many-body state.  When the probe layer is in a $\nu=1$ state, 
2D to 2D tunneling at temperature $T=0$ involves opposite spins for opposite directions of 
current flow, whereas when the probe layer is in a $\nu = 2$ state 2D to 2D tunneling involves both spins for current flow away from the probe layer, and is completely suppressed for the opposite bias.
The shared property of these two states that is responsible for their simplicity and for 
their effectiveness as tunneling probes is that they are single Slater determinant states with no correlations.  
With $\nu = 1$ and $\nu = 2$ as probes, we can directly use the sum-rules obtained above to estimate correlation energies 
by performing the series of measurements enumerated below.  This procedure is applicable for any target FQH state.  
A similar bias controlled spin selective tunneling technique 
was effectively employed some time ago  to study spin polarization in 
ferromagnets~\cite{Meservey1974} by taking advantage of spin-splitting of the 
BCS density-of-states peak in a superconductor.  

We remark that the use of filled Landau level integer quantum Hall states as tunnel probes
has potential limitations~\cite{Dial2007, Dial2010}  because of the high in-plane resistivities at these 
filling factors. Typically, the tunnel current enters the 2DES flowing perpendicular to the plane through a tunnel barrier. 
The measurement of a steady state current then requires it to have a path to exit the 2DES. 
If the two contacts are not perfectly aligned in the perpendicular direction, the current needs to 
flow in-plane to drain out to a distant contact. 
If the in-plane conductivity of the device is small, the current cannot flow out through the plane of the 2DES.
Time-domain capacitance spectroscopy~\cite{Dial2007, Dial2010} provides a potential way around this limitation by 
abandoning steady-state measurement and instead using isolated electrodes to capacitively apply the tunnel voltage and 
detect the tunneled electrons.

The chemical potentials, and zero temperature spectral functions are known exactly at the probe filling factors.  For  
$\nu = 1-\delta \nu $, 
\begin{subequations}\label{eq:nu=1_m}
\begin{align}
     & \mu = 2\epsilon_0\, , \\
     & \tilde{A}^{+}_{\downarrow}(\epsilon,\nu = 1-\delta\nu) = \delta(\epsilon+2\epsilon_0)\, , \\
     & \tilde{A}^{-}_{\uparrow}(\epsilon,\nu = 1-\delta\nu) = \delta(\epsilon)\, ,
\end{align}
\end{subequations}
at filling $\nu = 1+\delta \nu$, 
\begin{subequations}\label{eq:nu=1_p}
\begin{align}
     & \mu = 0\, , \\
     & \tilde{A}^{+}_{\downarrow}(\epsilon,\nu = 1+\delta\nu) = \delta(\epsilon)\, , \\
     & \tilde{A}^{-}_{\uparrow}(\epsilon,\nu = 1+\delta\nu) = \delta(\epsilon-2\epsilon_0)\, ,
\end{align}
\end{subequations}
and at $\nu = 2-\delta\nu$ filling,
\begin{subequations}\label{eq:nu=2}
\begin{align}
    & \mu = 2\epsilon_0\, ,\\
    & \tilde{A}^{-}_{\downarrow}(\epsilon,\nu = 2-\delta\nu) = \delta(\epsilon)\, , \\
     & \tilde{A}^{-}_{\uparrow}(\epsilon,\nu = 2-\delta\nu) = \delta(\epsilon)\,  .
\end{align}
\end{subequations}
Other partial spectral functions vanish identically.  
Notice the different chemical potentials associated with these spectral functions.
Since $\epsilon_0$, the energy of filled Landau level, is negative one can verify that 
the electron removal and addition spectral functions are non-zero at negative and positive 
energies respectively.  Because spin-polarization is not complete at finite temperatures, the
$\nu = 1$ spectral functions become more complex~\cite{Kasner1996}
developing separate peaks associated with minority spin removal and 
majority spin addition. Even with these additional complications at finite temperature, 
the $\nu = 1$ spectral functions that we propose using as probes are 
simpler and better understood than those of other FQH states. 

\begin{figure}[!htb]
  \includegraphics[width=0.45\textwidth]{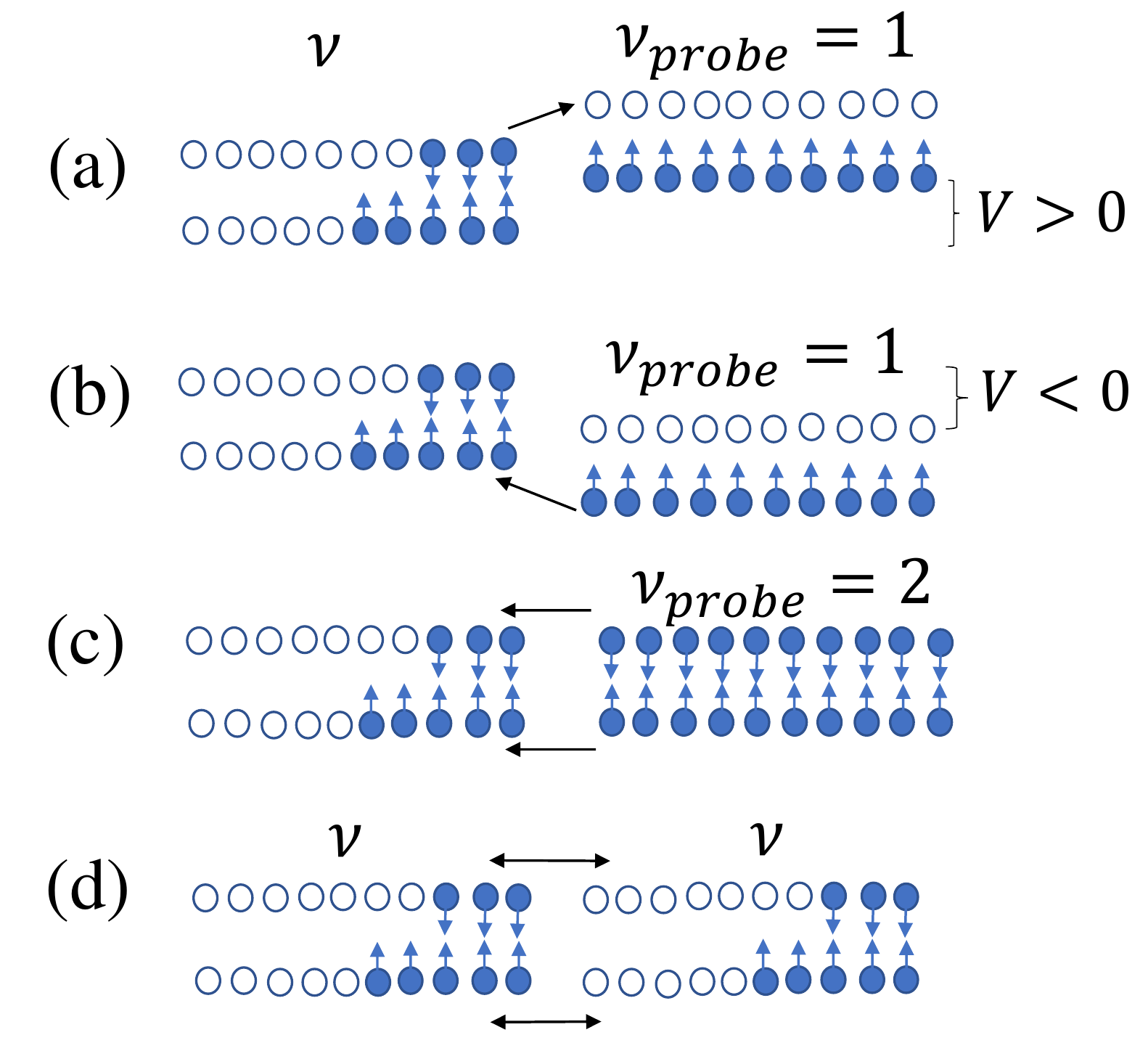}
  \caption{\label{Fig:Tunneling_experiment}
Schematic of proposed 2D to 2D tunneling experiments, (a) For a FQH state of interest at filling $\nu$ in one layer and a probe state at 
$\nu_{probe} = 1$ in the other layer positive bias drives electron flow of minority spin from layer $\nu$ to $\nu_{probe} = 1$, (b) 
At negative bias drives majority spin electron flow from  $\nu_{probe}$ to $\nu$. 
(c) When the probe layer has $\nu = 2$,  both spins flow out of the probe layer for one sign of bias and tunneling is suppressed for
the other sign of bias when the probe layer has a lower chemical potential.  
(d) When both layers are at filling $\nu$ and the state is partially spin-polarized both minority and 
majority spin electrons can contribute to tunneling for both signs of bias.}
\end{figure}

Consider then the following series of measurements.  
\begin{enumerate}

\item{\it 2D to 2D tunneling experiments with one layer at $\nu_{probe} = 1$ and 
the other layer at the filling factor being studied}:
(See Fig.~\ref{Fig:Tunneling_experiment}.(a), (b)). Since the
$\nu_{probe} = 1$ state is fully spin polarized this measurement
allows only minority spin tunneling to the $\nu_{probe} = 1$ layer, and only majority spin tunneling in the opposite direction. 
When the probe layer equilibrates at filling $\nu = 1-\delta \nu$ and is kept at a
higher bias only the down spin electrons flow from the target FQH layer to the probe layer.
From the $I-V$ data with this sign of bias (which we refer to as positive bias for convenience), 
one can directly obtain the electron removal portion of the 
spectral function of the target FQH layer using Eq.~\ref{Eq:IV_Convolution} and Eq.~\ref{eq:nu=1_m}: 
\begin{align}\label{Eq:Spectral_1_known}
      & \tilde{A}^{-}_{\downarrow}(\nu,eV) = \frac{I(-2\epsilon_0/e-V)}{I_0}\, .
\end{align}
Similarly, 
$\tilde{A}^{+}_{\uparrow}(\epsilon)$ can then be determined by bias in the opposite direction (negative bias) : 
\begin{align}\label{Eq:Spectral_2_known}
    & \tilde{A}^{+}_{\uparrow}(\nu,eV) = \frac{I(\epsilon_0/e)}{I_0}\, .
\end{align} 
It then follows from Eq.~\ref{Eq:Integral_IV} that the spin-dependent filling factors of the layer 
being studied satisfy
\begin{align}
      & \frac{\nu_{\downarrow}}{\nu_{\uparrow}} = \frac{\int^{\infty}_0 dV I(V)}{\int^{0}_{-\infty} dV I(V)}\, .
\label{eq:ratio}      
\end{align}
Since the total filling factor $\nu= \nu_{\uparrow} + \nu_{\downarrow}$ of the probe layer 
is normally known, Eq.~\ref{eq:ratio} 
allows the spin-dependent partial filling factors to be determined by this measurement along with the proportionality constant $I_0$.
In addition one can obtain four independent relations for the correlation energies relating to the experimental data.
From positive bias measurements we obtain
\begin{align}
     & \biggl (\frac{P^1}{P^0}\biggr )_{V > 0} = -2\epsilon_0\nu_{\downarrow} + \mu - \frac{2}{\nu_{\downarrow}}(\epsilon^c_{\downarrow\downarrow}+\epsilon^c_{\uparrow\downarrow})\, ,
\label{Eq:Corr_energy_down_probe}    
\end{align} 
and from negative bias measurements 
\begin{align}
     & \biggl (\frac{P^1}{P^0}\biggr )_{V < 0} = 2\epsilon_0\nu_{\uparrow} - \mu - \frac{2}{\bar{\nu}_{\uparrow}}(\epsilon^c_{\uparrow\uparrow}+\epsilon^c_{\uparrow\downarrow})\, .
\label{Eq:Corr_energy_up_probe}    
\end{align}
The subscripts $V>0, V<0$ above distinguish the moments obtained from positive or negative bias sector of $I-V$ data. 
All the quantities on the LHS of above expressions can be measured. 
Note that when the probe layer equilibrates at $\nu = 1+\delta\nu$ filling, the 
chemical potential shift across the incompressibility produces a corresponding 
shift in the moments.

\item {\it Probe layer at $\nu_{probe} = 2$ with other layer at target $\nu$}:
In an ideal system, the tunneling $I-V$ is strongly suppressed, when $\nu_{probe}$ is at the higher bias, 
since no states of either spin are available as tunneling final states. 
Non-zero tunneling occurs only when $\nu_{probe}$ is at a lower bias. The zeroth moment of the tunneling $I(V)$ satisfies,
\begin{align}\label{Eq:nu2_probe_0}
     & \sum_s P^0_s = I_0\sum_s \bar{\nu}_s\, ,
\end{align} 
which provides independent check of $I_0$. From the first moment of the tunneling $I(V)$, we obtain,
\begin{align}\label{Eq:nu2_probe}
      \sum_s P^1_s = I_0\sum_s[2\epsilon_0\nu_s\bar{\nu}_s-\mu\bar{\nu}_s-2(\epsilon^c_{s,s}+\epsilon^c_{s,-s})]\, .
\end{align}

\item Finally experiments can be performed with 
both layers at target filling $\nu$ (Fig.\ref{Fig:Tunneling_experiment} (c)). For some filling 
factors these experiments already exist in the literature and a specific case~\cite{Eisenstein2009,Eisenstein2016} was 
discussed in Sec.~\ref{Sec:Analysis}. Since, the individual spin currents are not resolved in $I-V$ plot, these experiment 
cannot accurately provide spin resolved moments $P^i_s$ in Eq.~\ref{Eq:Integral_IV}, \ref{Eq:current_moment_1st}. Instead can only accurately provide the total moment $P^i_\uparrow+P^i_\downarrow$. We obtain expression for correlation energies relating to the experimental data
\begin{align}
     &\sum_s P^1_s = -2I_0 \sum_s \big[ \epsilon^c_{s,s}+\epsilon^c_{s,-s} \big] \, .     
\label{Eq:Corr_energy_same_fill}     
\end{align}
\end{enumerate}

Eq.~\ref{Eq:Corr_energy_down_probe}, \ref{Eq:Corr_energy_up_probe}, and \ref{Eq:nu2_probe} determine the two 
correlation energies $\epsilon^c_{\uparrow\uparrow}+\epsilon^c_{\uparrow\downarrow}$ and $\epsilon^c_{\downarrow\downarrow}+\epsilon^c_{\uparrow\downarrow}$, and the chemical potentials $\mu$ of the target FQH state.
We re-emphasize that above equations are exact in the limit of no interlayer correlations.  
In the experiments with the probe layer at $\nu = 1$ or $\nu = 2$, the interlayer correlations are in general absent even if the layers are close. Eq.~\ref{Eq:Corr_energy_same_fill} obtained from tunneling between FQH layer at same filling provides an extra relation.
If the interlayer correlations are negligible for the tunneling between the same filling, the correlation energies obtained from the experiment with probe layer should satisfy Eq.~\ref{Eq:Corr_energy_same_fill}. This serves as a consistency check. However, any inconsistency between correlation energies obtained from first two experiment in the protocol (\textit{i.e.} Eq.~\ref{Eq:Corr_energy_down_probe}, \ref{Eq:Corr_energy_up_probe}, and \ref{Eq:nu2_probe}) and the third experiment (\textit{i.e.} Eq.~\ref{Eq:Corr_energy_same_fill}) implies the role of interlayer correlations. This way the above protocole can also qualitatively gauge the role of interlayer correlations.
We mention that, 
as described here, the spin dependent correlation energies can be determined, however, the contribution to spin dependent correlation energy from same spin and opposite spin correlations, \textit{i.e.} $\epsilon^c_{s,s}$  and $\epsilon_{s,-s}$ cannot be separated.


\section{Discussion \label{sec:discussion}}

We have calculated spin dependent spectral
 moment sum-rules for the TDOS in the FQH regime and related them to the measurements of 
 tunneling currents between FQH layers maintained, in the general case, at different filling factors. 
In so doing, 
we highlight the importance of equilibrium chemical potential differences between the two layers and the 
associated corrections to the sum-rules.
We show that for an arbitrary FQH state at filling factor $\nu$, 2D to 2D tunneling measurements
with partner probe layers at  $\nu = 1$ and $\nu = 2$
 can accurately determine spin-dependent partial filling factors, chemical potentials, and correlation energies.
These proposed tunneling experiments along with our spin dependent sum rules can 
potentially reveal more about the role of spin in general FQH states, which is in many cases not well understood.  
These experiments require 
tunneling between two FQH layers at different filling factors and therefore
require independent gate control of the two 
layers, as already employed in previous measurements of 
tunneling between $\nu = 5/2$ and $\nu = 7/2$~\cite{Eisenstein2017}.

The experiments by Eisenstein {\it et al.}~\cite{Eisenstein2009,Eisenstein2016} are likely in the regime 
in which interlayer correlations have a quantitative influence on the measured $I-V$ curves.  
Recently, the significant role of inter-layer excitonic effects was highlighted by Zhang {\it et al.}~\cite{Zhang2017} to explain the dependence of the peak in the $I-V$ curve 
on in-plane magnetic field. Chowdhury {\it et al.}~\cite{Chowdhury2018a} have argued
 that the puzzling behavior of small bias as function of in-plane magnetic is in the regime where the 
 charge spreading dynamics reflects the compressibility of composite Fermions. 
In the limit of small layer separations the bilayer system often forms an exciton condensate~\cite{Eisenstein2014} 
state in which inter-layer interactions drive broken symmetries.  
Our sum rules are exact in the opposite limit and should be 
applied to interpret experiments with large interlayer separations between layers 
and therefore weak inter-layer correlations. 
Detectable tunneling currents can be achieved at larger layer separations by 
reducing the height of the tunneling barrier.

Measurements of our sum-rules determine the actual correlation energies in 
real experimental systems,  not the theoretically calculated correlation energies of 
idealized model FQH states. The effects of finite layers thickness and Landau level mixing, 
when it can be described in terms of modified effective interactions, are 
accounted in correlation energies measured from tunneling $I(V)$'s in the way we describe.  
If there are small corrections related to interlayer tunneling, such that they are only additive contribution as shown in single mode approximation~\cite{Renn1994}, they simply renormalize the correlation energies obtained through our sum-rules. 


\textit{Acknowledgments}- We thank J. P. Eisenstein for sharing unpublished experimental results and for discussions 
which identified an important correction to an earlier version this manuscript. 
This work was supported by the Department of Energy, Office of Basic Energy Sciences under contract DE-FG02-02ER45958.  




\appendix
\section{Correlation energies and chemical potential\label{App:CE_expressions}}

Here we write down explicit expression for spin dependent correlation energies appearing in Eq.~\ref{Eq:Interaction_EOM_add}. The same spin correlation energy is

\begin{align}
     &\epsilon_{ss} = \frac{E_{ss}(\nu_s,\nu_{-s})}{N_{\phi}}\notag\\
     &\hspace{0.4cm}=\frac{1}{4}\langle \langle \sum_{n_1,n_2,n_3}
   (U^{n_1,n}_{n_2,n_3}-U^{n,n_1}_{n_2,n_3} )\notag\\
   &\hspace{2cm}\times (c^{\dagger}_{n_1 s}c^{\dagger}_{n s}c_{n_2s}c_{n_3s})\,\rangle\rangle\, . 
\label{Eq:SM_Correlation_same}  
\end{align}

While, the opposite spin correlation energy is

\begin{align}
     &\epsilon_{s,-s} = \frac{E_{s,-s}(\nu_s,\nu_{-s})}{N_{\phi}}\notag\\
     &\hspace{0.7cm} = \frac{1}{4}\langle \langle \sum_{n_1,n_2,n_3}
   (U^{n_1,n}_{n_2,n_3}-U^{n,n_1}_{n_2,n_3} )\notag\\
   &\hspace{2cm}\times (c^{\dagger}_{n_1, s}c^{\dagger}_{n, -s}c_{n_2,s}c_{n_3,-s})\,\rangle\rangle\, .
\label{Eq:SM_Correlation_opposite} 
\end{align}

And, the expression for filled Landau level energy per flux is

\begin{align}
   \epsilon_0 = \frac{1}{2} \sum_{n,n'} (U^{n,n'}_{n,n'}-U^{n,n'}_{n',n})\, .
\label{Eq:SM_LL_energy}  
\end{align}

Similarly, the chemical potentials 
\begin{align}
     &\mu = \frac{\partial E}{\partial N} = \sum_{\sigma,\sigma'}\frac{\partial}{\partial\nu} \epsilon_{\sigma,\sigma'}(\nu_{\uparrow},\nu_{\downarrow})\notag\\
     &\hspace{0.4cm} = \frac{\partial}{\partial\nu} \sum_{\sigma}\nu^2_{\sigma}\epsilon_0+\epsilon^c_{\sigma\sigma}+\epsilon^c_{\sigma,-\sigma}\, ,
\end{align}
in general depend on filling of both spin.

\section{Details on fitting\label{App:Fitting}} 
In this section we comment on our fitting procedure. We assume spectral functions can be approximated by a Gaussian. In total there are four Gaussians corresponding to electron addition and removal for both spins, which gives four fitting parameters. Tuning four fitting parameters can lead to over-fitting and not give unique best fit. So we systematically reduce number of free parameters here to get best fit. First, since $I-V$ plots are convolutions Eq.~\ref{Eq:IV_Convolution}, it reduces to three free parameters, i.e. $\sigma_s,\Delta_s,\nu_s$ for each spin,
\begin{align}
     &I_{s}(V) = \frac{I_0\nu_s\bar{\nu}_s}{\sqrt{2\pi}}\,\exp\biggl (-\frac{(V-\Delta_s)^2}{2\sigma^2_s}\biggr )\, .    
\label{Eq:IV_fit}
\end{align}
Here $\Delta_s = \mu_{s,+}-\mu_{s,-}$, and $\sigma_s = \sqrt{\sigma^2_{s,+}+\sigma^2_{s,-}}$ are obtained by convolution to two Gaussians associated with spectral functions. $\mu_{s,i}$ and $\sigma_{s,i}$ are peak position and standard deviation of spectral functions related to electron addition and removal.  We mention, $\mu_{s,+},\, \mu_{s,-},\,\sigma_{s,+},\,\sigma_{s,-}$ can not be determined separately from experiment. For, $\nu = 3/2$, since the two peaks are well separated in $I-V$ plot (Fig. \ref{Fig:Eisenstein_fit}), we first choose $\Delta_s$ at the bias corresponding to the two peak currents, i.e. $\Delta_{\uparrow} \sim 2.15\, mV,\, \Delta_{\downarrow} \sim 3.62\, mV$. Once peak values are adjusted, $\sigma_s$ are found using full width of half maxima of each Gaussian on the side away from the other peak, which gives, $\sigma_{\uparrow} \sim  0.56 mV,\, \sigma_{\downarrow} \sim 0.64 mV$. At last,
the relative peak height can be used to determine the factor in front of spin dependent $I-V$ curve to give
\begin{align}
    & \frac{I_{\uparrow}(V = 2.15\,mV)+I_{\downarrow}(V = 2.15 mV)}{I_{\uparrow}(V = 3.62\,mV)+I_{\downarrow}(V = 3.62 mV)} = \frac{0.238}{0.275}\, .  
\label{Eq:Peak_ratio}    
\end{align}
Assuming $\nu_{\uparrow} = 0.75+k$ and $\nu_{\downarrow} = 0.75-k$, above equation reads as,
\begin{widetext}
\begin{align}
    &\frac{1.786\,(-k^2-0.5k+0.1875)\,+\,0.11\,(-k^2+0.5k+0.1875)}{0.057\,(-k^2-0.5k+0.1875)\,+\,1.5625\,(-k^2+0.5k+0.1875)} = 0.865\, .
\label{Eq:Sup_filling_ratio}   
\end{align}
\end{widetext}
Which solving for $k$ gives, $k \sim 0.061$. Giving, $\nu_{\uparrow}\sim 0.81$, and $\nu_{\downarrow}\sim 0.69$. We can write down the spin resolved fit equation for tunneling current, and the fit equation,
\begin{align}
\begin{rcases}
    & I_{\uparrow} (V) = 0.219\,\exp\biggl (-\frac{[V-2.15]^2}{0.6262}\biggr )\, ,\notag\\
    & I_{\downarrow} (V) = 0.267\,\exp\biggl (-\frac{[V-3.62]^2}{0.8192}\biggr )\, .
\end{rcases}   \text{$\nu = 3/2$}    
\label{Eq:Fit_eq_1}
\end{align}
This procedure gives fairly good fit to the experimental $I-V$ curve.
Notice that the above fitting equation is not exactly the fit equation used in Fig.\ref{Fig:Eisenstein_fit} but still is very close to it. Starting from the above fit, we tune parameters slightly to make even better fit of Fig.\ref{Fig:Eisenstein_fit}, which is given by,
\begin{align}
\begin{rcases}
    & I_{\uparrow} (V) = 0.22\,\exp\biggl (-\frac{[V-2.05]^2}{0.57}\biggr )\, ,\notag\\
    & I_{\downarrow} (V) = 0.27\,\exp\biggl (-\frac{[V-3.66]^2}{0.83}\biggr )\, .
\end{rcases}   \text{$\nu = 3/2$}    
\end{align}
The above fit gives $k \sim 0.068$, with partial spin filling $\nu_{\uparrow} = 0.818,\, \nu_{\downarrow} = 0.682 $. Thus the fitting procedure is very accurate for$\nu = 3/2$ case.

For $\nu = 1/2$, if the majority spin filling is significantly large compared to the minority spin filling, the majority spin part gives most contribution to $I-V$ curve. This allows us to approximate no contribution from minority spin electron to the $I-V$ plot at bias smaller than the peak current bias. Following this argument, we first fix the majority spin contribution by matching the peak associated with it with the full $I-V$ peak and its broadening to the broadening towards the low bias side of $I-V$. This fixes the majority spin contribution to $I-V$ as,
\begin{align}
    & I_{\uparrow}(V) = 0.52\, \exp\biggl (-\frac{[V-2.81]^2}{1.75}\biggr )\, . 
\end{align}
Now, we subtract the area under $I_{\uparrow}-V$ curve from experimental $I-V$ curve, to find the total contribution (amplitude) due to minority spin electrons. This leads to the position of the peak maxima and the peak width for the minority spin current being the only adjustable parameters, which can be tuned to find the best overall fit.

\pagebreak

\bibliography{bibliography}


\end{document}